# Small-world structure of earthquake network

Sumiyoshi Abe[1] and Norikazu Suzuki[2]

[1]*Institute of Physics, University of Tsukuba, Ibaraki 305-8571, Japan*

[2]*College of Science and Technology, Nihon University, Chiba 274-8501, Japan*

Discoveries of the scale-free and small-world features are reported on a network constructed from the seismic data. It is shown that the connectivity distribution decays as a power law, and the value of the degrees of separation, i.e., the characteristic path length or the diameter, between two earthquakes (as the vertices) chosen at random takes a small value between 2 and 3. The clustering coefficient is also calculated and is found to be about 10 times larger than that in the case of the completely random network. These features highlight a novel aspect of seismicity as a complex phenomenon.

In modern statistical mechanics of complex systems, the concept of evolving random networks is attracting great interest. This stream was initiated by the pioneering works of Watts and Strogatz [1] on small-world networks and of Barabási and Albert on scale-free networks [2]. The primary purpose of this research subject is to understand topology and dynamics of evolving networks representing collective behaviors of complex systems. The worked examples include scientific collaboration, pattern of citation of scientific papers, metabolic networks in cells, food webs, social networks, World Wide Web, the Internet, electrical power grids, and collaboration of actors [1-4]. Each of these systems is mapped to a graph, in which a member (e.g., HTML of World Wide Web) is regarded as a vertex and the complex interaction between the members (e.g., pointing from one HTML to another through a link) is identified with an edge. Striking observations common in these apparently different systems are that their networks form small worlds with a few degrees of separation between two arbitrarily chosen members and are scale-free being characterized by the connectivity distributions that decay as a power law.

Seismology has also been attracting continuous interest of researchers of science of complex systems [5-13]. Although seismicity is characterized by extremely rich phenomenology, some of the known empirical laws are remarkably simple and exhibit its complexity aspects. The celebrated examples are the Omori law [14] for the temporal pattern of aftershocks showing slow relaxation and the Gutenberg-Richter law [15] for the relationship between frequency and moment obeying the power-law distribution.

In this article, we study the properties of the network associated with earthquakes. We report the discoveries of the scale-free and small-world features of the earthquake network constructed for southern California. The concept of earthquake network is the following. A geographical region under consideration is divided into a lot of small cubic cells. A cell is regarded as a vertex when earthquakes with any values of magnitude occurred therein. If two successive earthquakes occur in different cells, the corresponding two vertices are connected by an edge, whereas two successive earthquakes in the same cell define a loop. Complex fault-fault interactions are replaced by these edges and loops. In this way, the seismic data can be mapped to an evolving network. It is clear in this construction that there is a unique parameter, which is the size of the cell. Since there are no *a priori* operational rules to determine the cell size, it is essential to examine the dependencies of the network properties on this parameter. Once the cell size is fixed, the earthquake network is unambiguously defined by the seismic data [12]. The evolving earthquake network contains a few vertices corresponding to mainshocks. If the locus of a mainshock is shallow, it usually reorganizes the stress distribution in the relevant area, yielding the swarm of aftershocks. Through our data analysis, we have found a remarkable fact that aftershocks associated with a mainshock tend to return to the neighborhood of the locus of the mainshock, geographically, making the degree of connectivity of the mainshock vertex very large. Accordingly, the mainshock plays a role of a "hub" with large degrees of connectivity and preferential attachment [2] is realized. This is schematically depicted in Fig. 1, in which the vertices, *A* and *B*, may be identified with the mainshocks. This observation leads to the reasoning that the earthquake network may be scale-free and possess the small-world structure. Here, we show that this is indeed the case. We shall see that the connectivity distribution decays as a power law, the degrees of separation take the small value between 2 and 3, and furthermore the clustering coefficient [1] is about 10 times larger than that in the case of the completely random network [3,16]. These results highlight a novel aspect of seismicity as a complex phenomenon.

In the Barabási-Albert scale-free network model [2], a newly created vertex tends to be connected to the *i*th vertex $v_i$ having connectivity $k_i$ with probability

$$\Pi(k_i) = \frac{k_i + 1}{\sum_j (k_j + 1)}, \tag{1}$$

which mathematically represents the nature of preferential attachment. In the continuous and long-time limits, the model gives as the solution the connectivity distribution of the Zipf-Mandelbrot type [17]

$$P(k) \sim \frac{1}{(k + k_0)^\gamma}, \tag{2}$$

where $\gamma$ and $k_0$ are positive constants and $\gamma > 1$, in particular. In the worked examples [3], the exponent $\gamma$ ranges between 1.05 and 3.4.

Now, the data we analyzed are those made available by the Southern California Earthquake Data Center (http://www.scecdc.scec.org/catalogs.html). They are in the time interval between 00:40:07.47 on January 1, 1992 and 23:55:34.66 on December 31, 1992, covering the region 29°15.25'N–39°35.21'N latitude, 114°14.14'W–121°45.85'W longitude, and 0.00km-57.88km in depth (of the foci of the observed earthquakes). The total number of earthquakes with arbitrary values of magnitude in this spacetime region is 51416.

In Fig. 2, we present the plot of the connectivity distribution. The cell size employed here is $5\text{km} \times 5\text{km} \times 5\text{km}$. The dots represent the data in the above-mentioned region, whereas the solid line describes the Barabási-Albert model. From it, we can appreciate that the data well obey the distribution in Eq. (2). We have also examined other values of the cell size, but the result turned out to always exhibit the scale-free nature. This may be interpreted as follows. The Gutenberg-Richter law, on the one hand, tells us that frequency of earthquakes with large values of moment decays as a power law. On the other hand, as already mentioned, aftershocks associated with a mainshock tend to be connected to the vertex of the mainshock. Thus, the scale-free nature of the connectivity distribution is consistent with the Gutenberg-Richter law.

Next, we show in Fig. 3 the degrees of separation (i.e., the characteristic path length or the diameter) between an arbitrary pair of two vertices. Here, we vary the cell size from $5 \text{km} \times 5 \text{km} \times 5 \text{km}$ to $10 \text{km} \times 10 \text{km} \times 10 \text{km}$ by every $1 \text{km}$. The values of the degrees of separation were calculated by random sampling of 60 pairs of vertices. The degrees tend to slightly decrease with respect to the cell size, as it should do. The values are typically in-between 2 and 3, very small, showing the small-world nature of the earthquake network.

Finally, we discuss the clustering coefficient, $C$, proposed by Watts and Strogatz in Ref. [1]. The scale-free nature of the connectivity distribution leads to the expectation that the value of $C$ is much larger than that in the case of the completely random network whose connectivity distribution is Poissonian [2,3,16]. To calculate $C$, it is important to notice that loops attached to a single vertex should be removed and multiple edges between two vertices have to be identified with a single edge. For example, consider three vertices, $v_1$, $v_2$, and $v_3$. Suppose they are originally connected as $v_1 \to v_2 \to v_1 \to v_2 \to v_2 \to v_3$. This should be identified with $v_1 \to v_2 \to v_3$ now.

The definition of the clustering coefficient is the following. Assume that the $i$th vertex $v_i$ has $k_i - 1$ neighboring vertices. At most, $k_i(k_i - 1)/2$ edges can exist between them. Calculate $c_i \equiv$ (number of edges of $v_i$ and its neighbors) $/ [k_i(k_i - 1)/2]$. Then, the clustering coefficient is defined by

$$C = \frac{1}{N} \sum_{i=1}^{N} c_i, \qquad (3)$$

where $N$ denotes the total number of vertices. In the case of the completely random network, this quantity can be written as follows [1,3,4]:

$$C = C_{random} = \frac{<k>}{N} << 1, \qquad (4)$$

where $<k>$ is the average connectivity. The point of central importance is that a complex network has the clustering coefficient which is much larger than $C_{random}$ [1].

We have analyzed two subintervals: (I) between 01:50:15.81 on January 30, 1992 and 05:48:10.93 on February 2, 1992, with 63 events, and (II) between 11:57:34.10 on June 28, 1992 and 20:48:22.61 on June 28, 1992, with 176 events. In the period (I), seismicity is moderate, whereas it is very active in the period (II). In particular, the initial time of the period (II) is adjusted to be the event of the mainshock with M7.3 (34°12.01'N latitude, 116°26.20'W longitude, and 0.97km in depth), followed by the swarm of aftershocks. This is why the period (II) is much shorter than (I). The cell size is taken to be 10km×10km×10km for (I) and (II). Both of the corresponding earthquake networks have 50 vertices. The results are: (I) $C_{actual} = 0.680$ ($C_{random} = 0.046$), (II) $C_{actual} = 0.653$ ($C_{random} = 0.093$). Therefore, compared to the completely random network, the clustering coefficient is about 10 times larger.

In conclusion, we have constructed the complex network for earthquakes and have studied its scale-free and small-world features. We have shown that the connectivity distribution decays as a power law and the degrees of separation between two earthquakes as the vertices chosen at random take a small value between 2 and 3. We have also calculated the clustering coefficient and have found that its actual value is about 10 times larger than that in the case of the completely random network.

S. A. was supported in part by the Grant-in-Aid for Scientific Research of Japan Society for the Promotion of Science.

# Figure Captions

Fig. 1  A schematic description of the earthquake network. The vertices represent the earthquakes and the edges replace complex fault-fault interactions. *A* and *B* are mainshocks and have large connectivities.

Fig. 2  Plot of the connectivity distribution. The solid line corresponds to the model in Eq. (2) with $\gamma = 1.61(\pm 0.03)$ and $k_0 = 1.27(\pm 0.02)$. All quantities are dimensionless.

Fig. 3  The degrees of separation for various cell sizes. The values are $2.87(\pm 0.47)$ for $5\text{km} \times 5\text{km} \times 5\text{km}$ ($N = 3434$), $2.80(\pm 0.51)$ for $6\text{km} \times 6\text{km} \times 6\text{km}$ ($N = 2731$), $2.58(\pm 0.56)$ for $7\text{km} \times 7\text{km} \times 7\text{km}$ ($N = 2165$), $2.63(\pm 0.49)$ for $8\text{km} \times 8\text{km} \times 8\text{km}$ ($N = 1827$), $2.65(\pm 0.58)$ for $9\text{km} \times 9\text{km} \times 9\text{km}$ ($N = 1582$), and $2.53(\pm 0.54)$ for $10\text{km} \times 10\text{km} \times 10\text{km}$ ($N = 1344$), where *N* denotes the number of the vertices. The error bounds are given by the standard deviations for 60 pairs of vertices chosen at random.

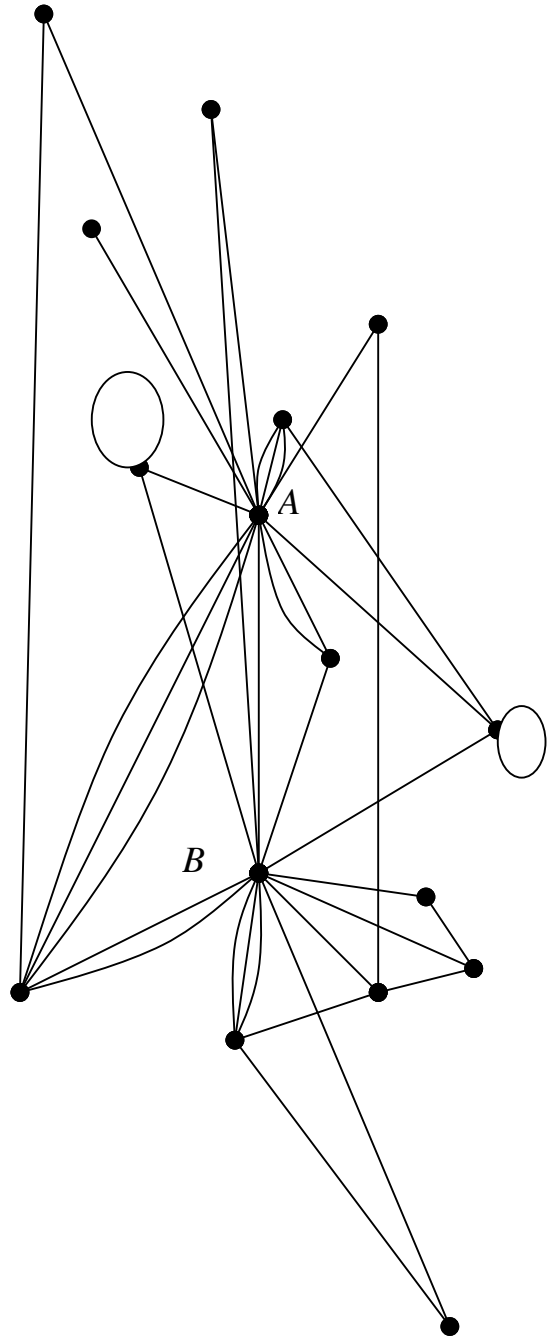

Fig.1

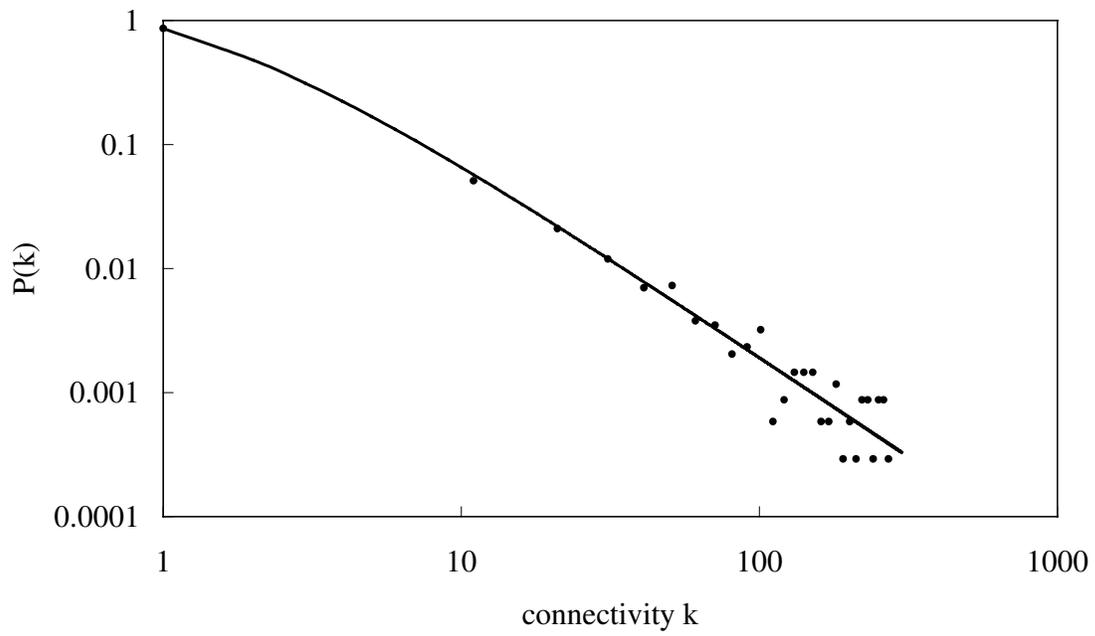

Fig. 2

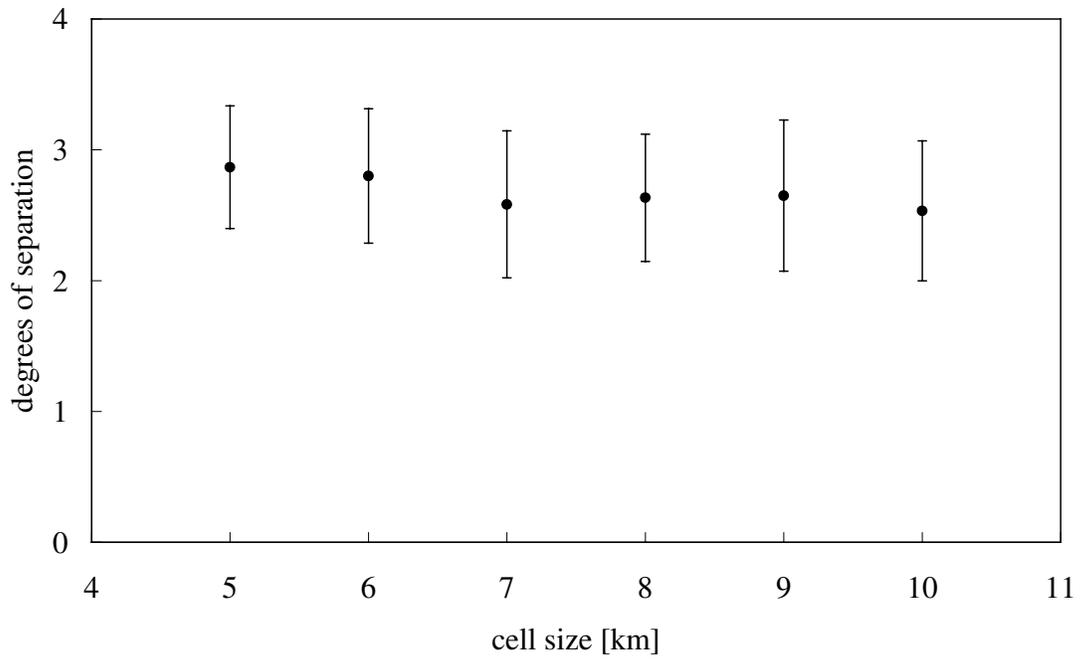

Fig. 3